\documentclass[aps,pre,twocolumn,showpacs]{revtex4-1}
\usepackage[colorlinks,linkcolor=blue,citecolor=blue,urlcolor=blue]{hyperref}
\usepackage{epsfig}
\usepackage{amsmath}
\usepackage{times}
\usepackage{graphicx}% Include figure files
\usepackage{dcolumn}% Align table columns on decimal point

\begin{document}

\title{Behavior of susceptible-vaccinated--infected--recovered epidemics with diversity in the infection rate of the individuals}
\author{Chao-Ran Cai}
\author{Zhi-Xi Wu}\email{wuzhx@lzu.edu.cn}%{Corresponding author: wuzhx@lzu.edu.cn}
\author{Jian-Yue Guan}\email{guanjy@lzu.edu.cn}
\affiliation{Institute of Computational Physics and Complex Systems, Lanzhou University, Lanzhou, Gansu 730000, China}

%\date{\today}
\begin{abstract}
We study a susceptible-vaccinated--infected--recovered (SVIR) epidemic-spreading model with diversity of infection rate of the individuals. By means of analytical arguments as well as extensive computer simulations, we demonstrate that the heterogeneity in infection rate can either impede or accelerate the epidemic spreading, which depends on the amount of vaccinated individuals introduced in the population as well as the contact pattern among the individuals. Remarkably, as long as the individuals with different capability of acquiring the disease interact with unequal frequency, there always exist a cross point for the fraction of vaccinated, below which the diversity of infection rate hinders the epidemic spreading and above which expedites it. The overall results are robust to the SVIR dynamics defined on different population models; the possible applications of the results are discussed.
\end{abstract}
%\pacs{02.50.Le, 87.23.Ge, 05.65.+b, 64.75.+g}
\maketitle

\section{Introduction}\label{intro}
Infectious diseases have always been the great enemy of human health. Historically, large outbreaks of epidemic usually posed a great threat to health and caused great loss for individuals. In some sense, the history of humans is a history of struggle with all kinds of diseases, from the Black Death in medieval Europe to the recently notorious severe acute respiratory syndrome~\cite{SARS1,SARS2,SARS3}, avian influenza~\cite{ai1,ai2}, swine influenza~\cite{h1n11,h1n12}, etc.

So far, vaccination is the most effective approach to preventing transmission of vaccine-preventable diseases, such as seasonal influenza and influenza like
epidemics, as well as reducing morbidity and mortality~\cite{Bauch:1}. In a voluntary vaccination program, the individuals are subject not only to social factors such as religious belief and human rights, but also to various other conditions such as risk of infection, prevalence of disease, coverage, and cost of vaccination.

Recently, a great deal of effort has been devoted to the investigation of the interplay between vaccine coverage, disease prevalence, and the vaccinating behavior of individuals by integrating game theory into traditional epidemiological models \cite{Bauch:1,Bauch:2,Vardavas:1,Brehan2007pre,haifeng,fufeng,Brehan2007pre, Perra2011plos,Zhang2012pa,Liu2012pre,Peng2013pre,Zhang2013pre,Cardillo2013pre}. For brief reviews of this research topic, we refer the reader to Refs.~\cite{Funk2010jrsi,Bauch2013science} and reference therein. Bauch $et\ al.$ used game theory to explain the relationship between group interest and self-interest in smallpox vaccination policy~\cite{Bauch:1,Bauch:2} and found that voluntary vaccination was unlikely to reach the group-optimal level. Vardavas and co-workers investigated the effect of voluntary vaccination on the prevalence of influenza based on minority game theory and showed that severe epidemics could not be prevented unless vaccination programs offer incentives~\cite{Vardavas:1}. Zhang $et\ al.$ studied the epidemic spreading with voluntary vaccination strategy on both Erd\"{o}s-R\'{e}nyi random graphs and Barab\'{a}si-Albert scale-free networks~\cite{haifeng}. They found that disease outbreak can be more effectively inhibited on scale-free networks rather than on random networks, which is attributed to the fact that the hub nodes of scale-free networks are more inclined to getting vaccinated after balancing the pros and cons. More recently, Fu and co-workers proposed a game-theoretic model to study the dynamics of vaccination behavior on lattice and complex networks~\cite{fufeng,Zhang2012pa}. They found that the population structure causes both advantages and problems for public health: It can promote voluntary vaccination to high levels required for herd immunity when the cost for vaccination is sufficiently small, whereas small increases in the cost beyond a certain threshold will cause vaccination to plummet, and infection to rise, more dramatically than in well-mixed populations. Another research line studying the effect of human behavior on the dynamics of epidemic spreading considers mainly the coevolution of node dynamics and network structure (the so-called adaptive networks~\cite{Gross2006prl}), which can affect considerably the spreading of a disease~\cite{Shaw2010pre,Nuno2012jsm,Wang2011jpa}.

In most classical epidemiological models~\cite{Anderson1991book,Keeling2008book}, the individuals in the population are assumed to be identical, e.g., all susceptible individuals acquire the disease with the same probability whenever in contact with an infected individual, and all infected individuals recover, or go back to being susceptible, with the same rate. Such consideration is, however, far from the actual situation. Generally, catching a disease could be caused by many complex factors and there might be great difference among the individuals in the contact rate~\cite{Volz2008epjb}, the infection rate (or disease transmission rate)~\cite{Nuno2012jsmb,Buono2013pre}, the recovery rate, the cost when the individual is infected, and so forth. One example of such a scenario would be the case where the population is divided into a relatively wealthy class (e.g., representing urban residents), which is less susceptible to infectious disease being considered due to better living conditions and/or health care, and a class of relatively impoverished (e.g., representing rural residents), which is more susceptible to infection. An alternative view is to regard roughly the whole population as composed of two main groups, say, youths and adults, where the former is more resistant to disease than the latter, owning to their stronger physique and immune system.

In the present work, we relax the assumption of identical nature of the individuals and take into account their heterogeneity in acquiring disease when in contact with infectious individuals. To do this, we divide the whole population into two groups, youths (hereafter group $A$) and adults (group $B$) for simplicity, with the same size, and assume that the individuals from group $B$ are more likely to be infected than those from group $A$. For the sake of comparison, we presume that only the disease transmission rate for the individuals in the two groups are distinct and other parameters, including the recovery rate, the cost of infection, and the cost for vaccination, are identical. By doing so we hope to catch any possible effects on disease prevalence and vaccination coverage caused by the variability of susceptibility. Our results presented below show that the heterogeneity in infection rate has a significant influence on disease spreading and hence cannot be ignored in the forecast of epidemic size and vaccination coverage.

Our paper is organized as follows. In Sec.~\ref{model} we define our model and give detailed information for the numerical simulation method and the parametrizations. In Sec.~\ref{results} we present and analyze the main results of our model. We summarize and discuss the results in Sec.~\ref{conclusion}.

\section{Model definition}\label{model}

\begin{figure*}
\begin{align*}
\epsfig{figure=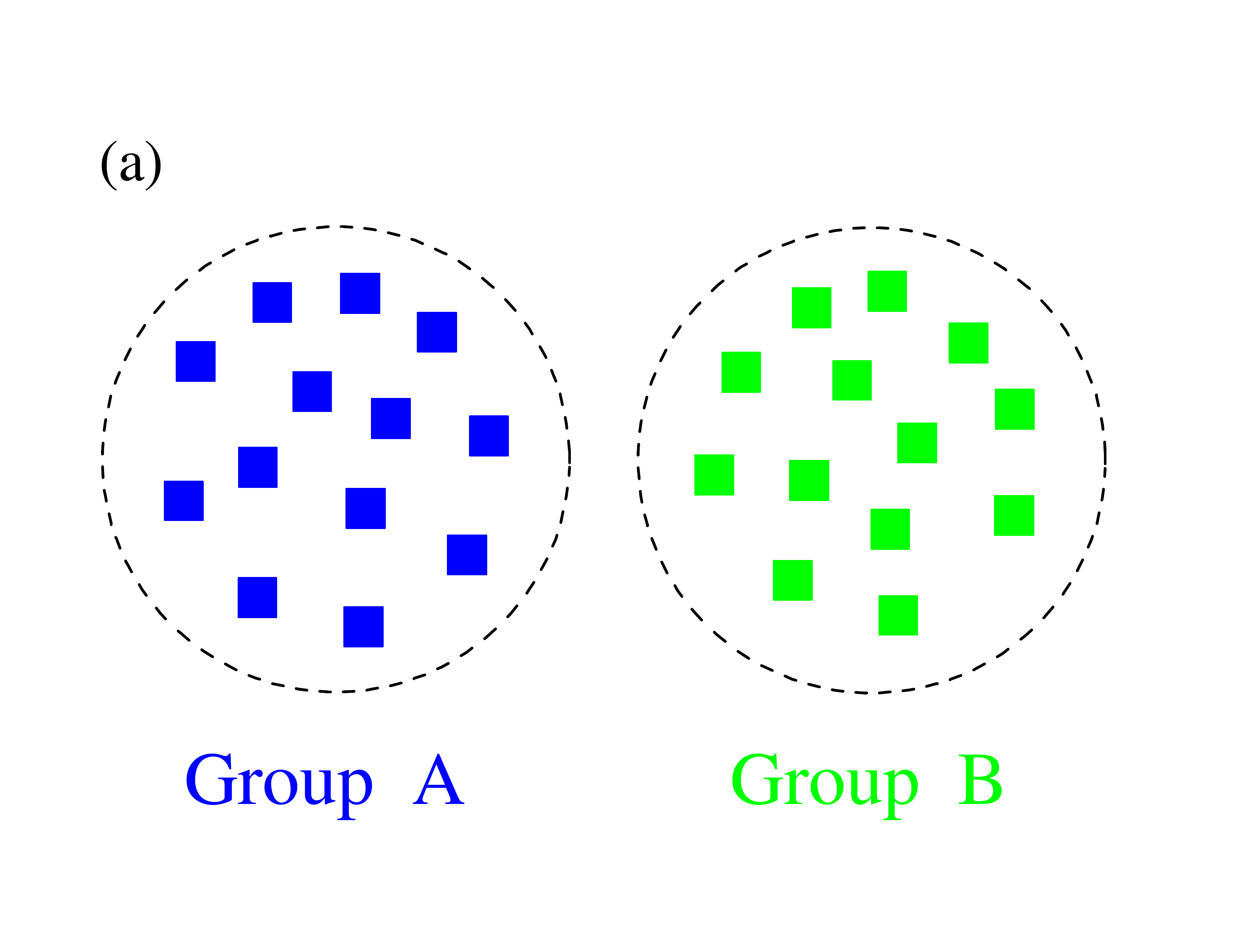,width=0.3\linewidth}
\epsfig{figure=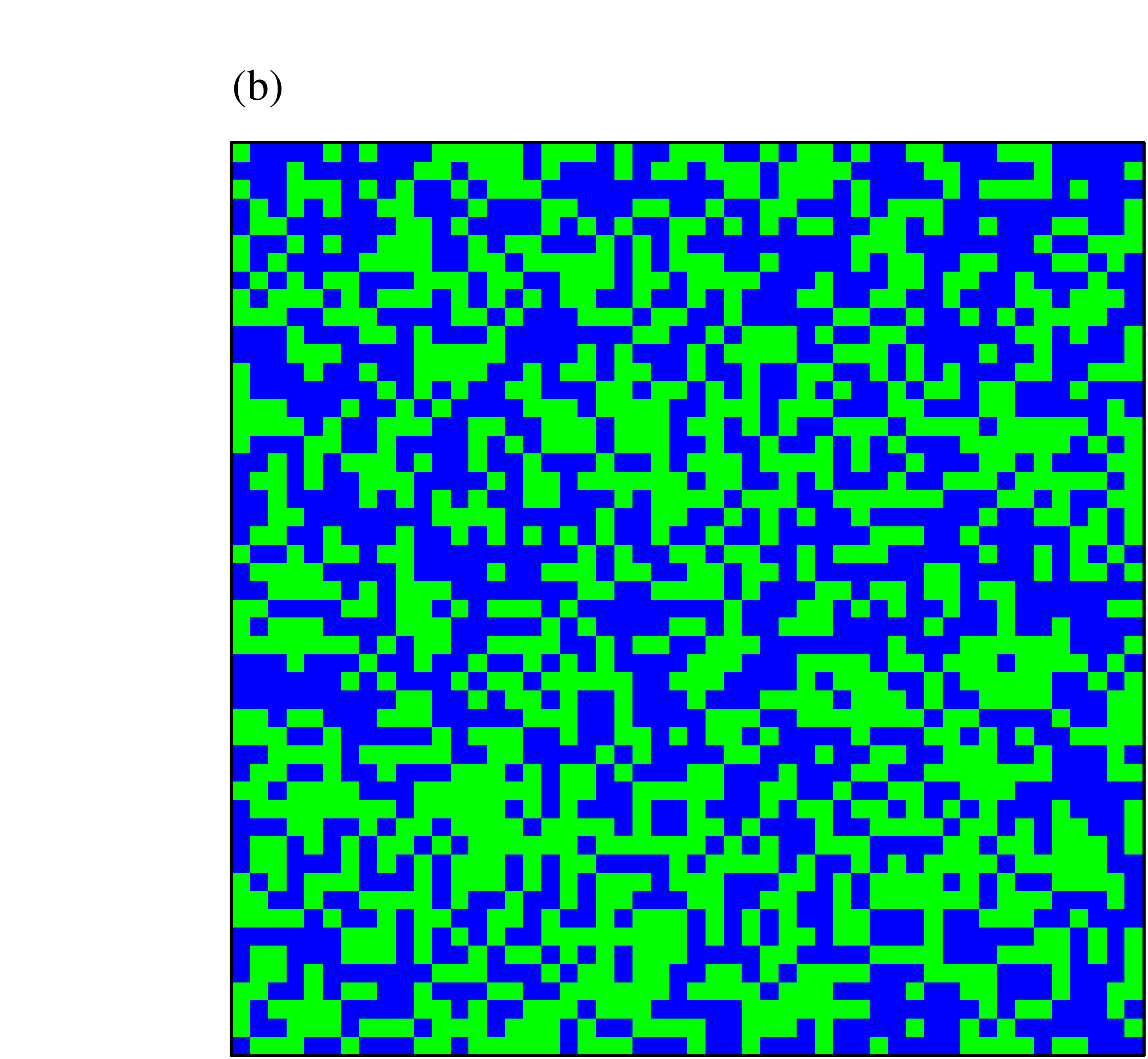,width=0.3\linewidth}
\epsfig{figure=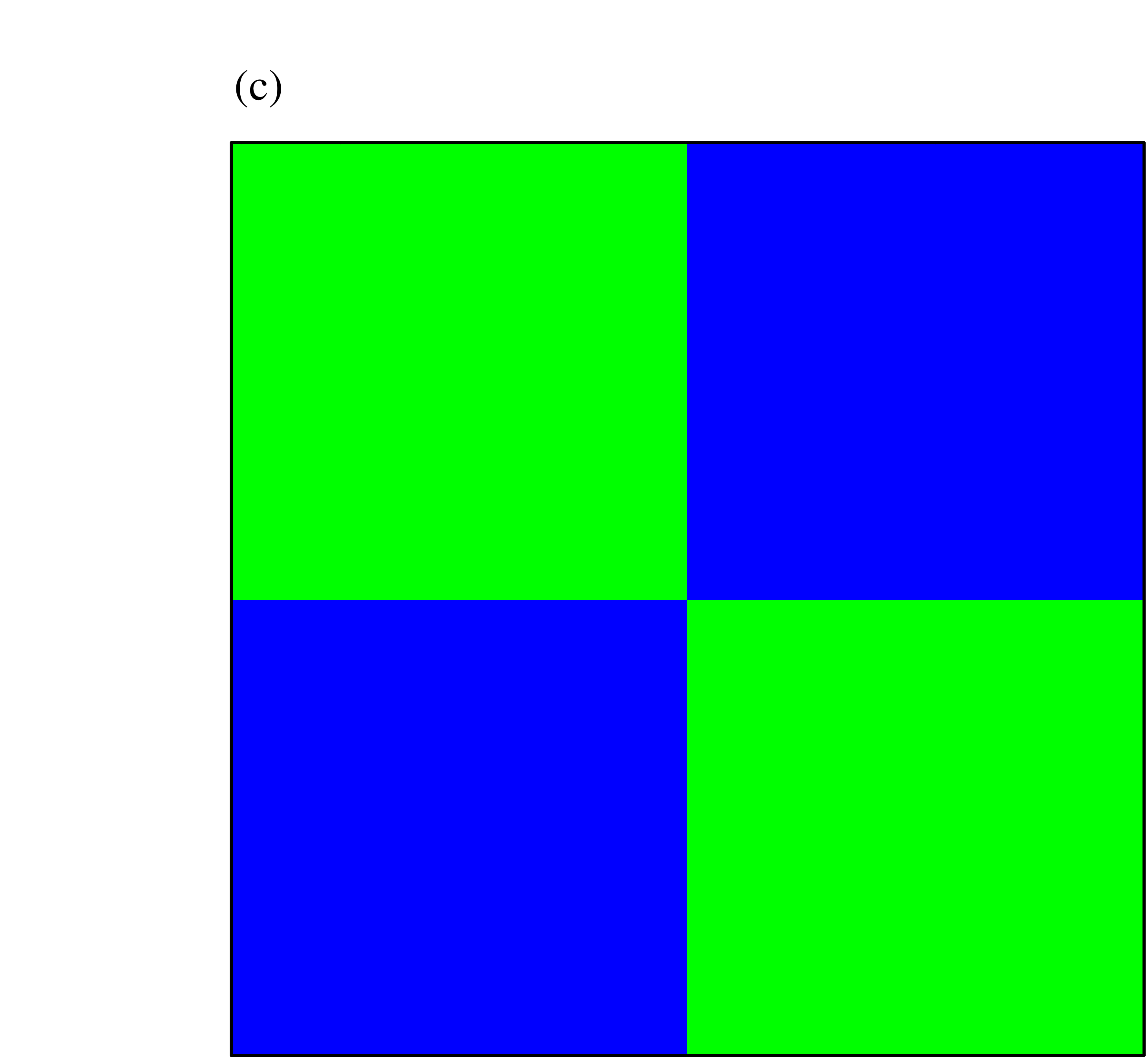,width=0.3\linewidth}
\end{align*}
\caption{(Color online) Schematic illustration of population models we studied in the main text. (a) Simple metapopulation model composed of two subpopulations, within each one consisting of the same type of individuals. (b) Individuals from different groups are \emph{randomly} arranged on a square lattice. (c) Individuals from different groups are \emph{regularly} arranged on the lattice. The $A$-type and $B$-type individuals are indicated by blue (dark gray) and green (light gray) squares, respectively.}\label{Fig:lattice}
\end{figure*}

It is well known that the contact pattern among individuals dramatically impacts the spatiotemporal dynamics of epidemic spreading in a population~\cite{Anderson1991book,Keeling2008book}. In order to examine the robustness of the results of our model, we consider two types of population models, namely, a simple metapopulation model and a spatially structured population model, as illustrated in Fig.~\ref{Fig:lattice}.

In the metapopulation model, the whole population is divided into two subpopulations with equal size, namely, group $A$ and group $B$. Within each subpopulation, the individuals are assumed to be homogeneously mixed, that is, every individual has the same opportunity to be in contact with everyone else. Generally speaking, because of the diversity in social conditions or lifestyles, the individuals living in an urban area would be more likely to interact with those also living the same area and less likely to interact with those in the suburb. Therefore, we consider the distinct contact pattern among the individuals to study its impact. This is done by assuming that any pair of individuals from different (the same) groups have an interaction frequency $\epsilon$ (1-$\epsilon$). Here $\epsilon$ is restricted to the interval [0,0.5]. In the spatially structured population model, we consider two kinds of occupation of the individuals on a square lattice to introduce the diversity of interaction pattern among them. To be more specific, in the first case, the youths and the adults are arranged in a random way such that they can interact with the same frequency, which is similar to the case of $\epsilon=0.5$ in the metapopulation case. In the second case, the individuals are regularly prearranged to gather together with the same type of individuals [see Fig.~\ref{Fig:lattice}(c)]. In this way, we are able to investigate how the mixing pattern affects the epidemic spreading in the population.

We implement our susceptible-vaccinated--infected--recovered epidemic-spreading dynamics in the following way. The epidemic strain infects an initial number of individuals $I_0$ and then spreads in the population according to the classical susceptible-infected-recovered(SIR) epidemiological model, with per-day transmission rate $r$ for each pair of susceptible-infected contact and recovery rate $g$ for each infected individual getting immune to the disease. Whenever the vaccinated compartment is involved in the epidemiological model, a fraction $f_V$ of individuals are randomly chosen in the whole population in the initial stage to get vaccinated. For simplicity, here we assume that vaccination grants perfect immunity for the infectious disease. The epidemic continues until there are no more newly infected individuals. As such, those unvaccinated susceptible individuals would either be infected or successfully escape from infection at the end of each spreading season.

In realistic situations, to vaccinate or not to vaccinate is sometimes the business of the individuals. Thus, except for the above case where the fraction of vaccinated individuals is compulsively introduced, we also consider a voluntary vaccination program for preventing an influenzalike infectious disease, in which individuals need to decide whether or not to receive a vaccine each season based on their perceived risk of disease infection. Following previous studies \cite{fufeng,Vardavas:1,Brehan2007pre,Zhang2012pa}, we model the vaccination dynamics as a two-stage game. At the first stage, each individual decides whether or not to get vaccinated, which will incur a cost $C_V$, including the immediate monetary cost for vaccine and the potential risk of vaccine side effects. Individuals catching the epidemic will suffer from an infection cost $C_I$, which may account for disease complications, expenses for treatment, etc. Those individuals who escape infection are free riders and pay for nothing. Without loss of generality, we set $C_I=1$ and let $c=C_V/C_I$ describe the relative cost of vaccination, whose value is restricted in the region of [0,1] (otherwise, doing nothing would be better than getting vaccinated). The second stage is the same epidemic spreading processes as described before. After each spreading season, the individuals are allowed to rechoose their choice for vaccination based on a pairwise comparison rule (more details will be given below).

We carry out stochastic simulations for the above epidemiological (game-theoretic) processes in both population models, wherein each seasonal epidemiological process is implemented by using the well-known Gillespie algorithm~\cite{Gillespie1976403,Gillespie}. In particular, at any time $t$, we calculate each individual's transition rate $\lambda_i(t)$. The rate for any susceptible individual becoming infected is $\lambda_i(t)=r\times k_{\textrm{inf}}$ and $k_{\textrm{inf}}$ is the number of infected neighbors of the focal individual. The rate for any infected individual recovering is $\lambda_i(t)=g$. The recovered individuals do not change state and the rate for them is therefore $\lambda_i(t)=0$. Summing up all of them, we yield the total transition rate $\omega(t)=\Sigma_i \lambda_i(t)$. With this value in hand, the time at which the next transition event occurs is $t'=t+\Delta t$, where $\Delta t$ is sampled from an exponential distribution with mean $\frac{1}{\omega(t)}$ (if we generate a uniform random number $u\in [0,1)$,  then the time interval is $\Delta t=-\frac{\ln(1-u)}{\omega(t)}$). The individual whose state is chosen to change at time $t'$ is sampled with a probability proportional to $\lambda_i(t)$. That is, a uniform random number $v\in [0,1)$ is generated and if $\Sigma_{j=1}^{k-1}\lambda_j(t)/ \omega (t)< v <\Sigma_{j=1}^{k}\lambda_j(t)/ \omega (t)$, then individual $k$ is chosen to change state. This elementary step is repeated until there are no infected individuals left in the population.

\section{Analysis and Results}~\label{results}

\subsection{Metapopulation without vaccinated compartment}
We first examine our model in metapopulations. For convenience, the two groups $A$ and $B$ are denoted by the subscripts $a$ and $b$, respectively. According to the above illustrated scenario, the time evolution of population states for group $A$ can be expressed as the following deterministic ordinary differential equations:
\begin{eqnarray}
&\frac{dS_a}{dt}=-r_a NS_a[(1-\epsilon)I_a+\epsilon I_b]\label{dS_xdt},\\
&\frac{dI_a}{dt}=r_a NS_a[(1-\epsilon)I_a+\epsilon I_b]-gI_a\label{dI_xdt},\\
&\frac{dR_a}{dt}=gI_a\label{dR_xdt}.
\end{eqnarray}
As mentioned before, the parameter $\epsilon$ is the cross contact coefficient, which stands for the contact frequency between individuals from different groups.

For the whole system that includes groups $A$ and $B$, we have the following equations:
\begin{eqnarray}
\nonumber&\frac{dS}{dt}=-r_aN S_a\left[\left(1-\epsilon\right)I_a+\epsilon I_b\right]
-r_bN S_b\left[\left(1-\epsilon\right)I_b+\epsilon I_a\right]\label{dSdt},\\
&\\
\nonumber&\frac{dI}{dt}=r_aN S_a\left[\left(1-\epsilon\right)I_a+\epsilon I_b\right]\\
&+r_bN S_b\left[\left(1-\epsilon\right)I_b+\epsilon I_a\right]-gI\label{dIdt},\\
&\frac{dR}{dt}=g(I_a+I_b)=gI\label{dRdt}.
\end{eqnarray}
In the limit $\epsilon\rightarrow0$, the basic reproduction number (whose value identifies the expected number of secondary infections produced by an infected individual during that individual's infectious period within the entire susceptible population) of groups $A$ and $B$ can be approximately written as $R_{0a}=r_aN/g$ and $R_{0b}=r_bN/g$, respectively. By taking the average over each group, we obtain the \emph{effective} basic reproduction number of the infectious disease $\mathcal{R}_0=(r_a+r_b)N/2g=\langle r\rangle N/g$~\footnote{Note that only in the limit case of $\epsilon\rightarrow0$ can $\mathcal{R}_0$ be approximately written as $(r_a+r_b)N/2g$. In any cases $\epsilon\gg0$, we are unable to write out the explicit form of $\mathcal{R}_0$, but just keep the quantity $\protect\langle r\protect\rangle=(r_a+r_b)/2$ as constant.} where $\langle r\rangle$ is the average value of the disease transmission rate of the whole population.

By varying the value of $r_a$ and $r_b$, we are able to introduce the difference in transmission rate of the infectious disease for the individuals. For the sake of comparison, we keep the average value of the transmission rate fixed as $\langle r\rangle=(r_a+r_b)/2$. Denoting $r_a/r_b$ by $x$, the relative disease transmission rate for the two types of individuals, after some simple algebra we have
\begin{equation}\label{r_a}
r_a=\frac{2x\langle r\rangle}{1+x}, \quad\mathrm{and}\quad
r_b=\frac{2\langle r\rangle}{1+x}.
\end{equation}
When $x$ is close to zero, there exists a great difference between the individuals in group $A$ and those in group $B$ in acquiring the disease (i.e., we consider the case where the youths are very resistant to the infection, while the adults are very vulnerable to the disease). As $x$ goes to unity, the variation of the disease transmission rate among the two groups vanishes.

\begin{figure}
\epsfig{figure=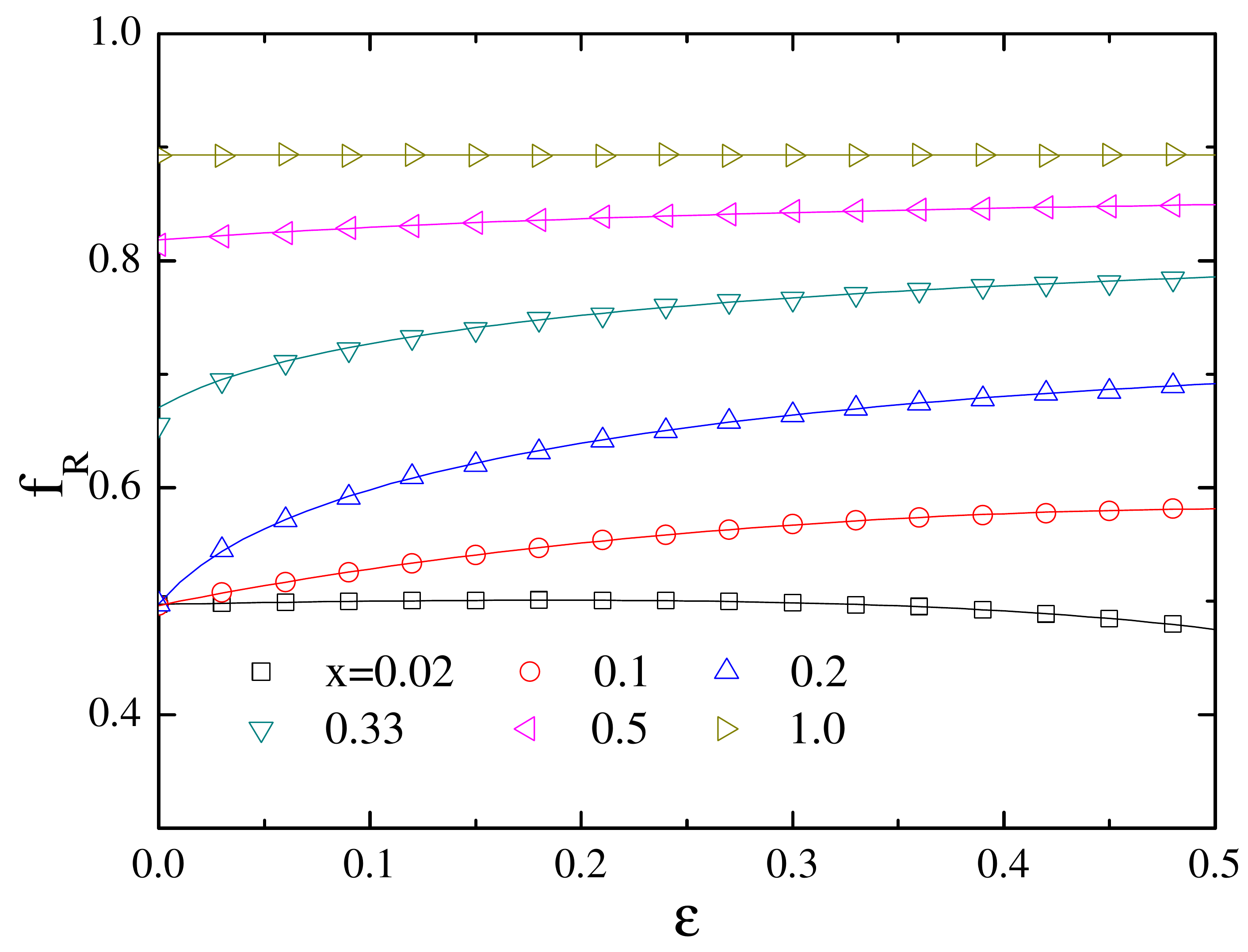,width=\linewidth}
\caption{(Color online) Epidemic spreading in the metapopulation model without the vaccinated compartment. The final epidemic size $f_R$ is plotted as a function of the cross coefficient $\epsilon$ for several different values of the relative disease transmission rate $x$. The lines are for the analytical predictions from Eqs.~(\ref{dSdt})--(\ref{dRdt}). The symbols are simulations obtained by carrying out the Gillespie algorithm. The parameters are the total population size $N=N_A+N_B=10\ 000$, average value of the disease transmission rate $\langle r\rangle=\frac{2.5}{3N}$ day$^{-1}$ person$^{-1}$, recovery rate $g=\frac{1}{3}$ day$^{-1}$, and number of initial infection seeds $I_a$=$I_b$=10. Simulation results are averaged over 100 independent runs.}
\label{Fig:meta-without}
\end{figure}

Let us show in Fig.~\ref{Fig:meta-without} the influence of the cross contact coefficient $\epsilon$ on the epidemic spreading in the population without a vaccinated compartment. In the case of the limit $x\rightarrow1$, we have $r_a\approx r_b$, which means that the possibilities of acquiring the disease through susceptible-infected contact for the individuals from the two groups are almost the same. As a consequence, the final epidemic size $f_R$, i.e., the average fraction of recovered individuals in the whole population, does not change much as the parameter $\epsilon$ varies. Note that with the current parametrization settings the final epidemic size without vaccination is about $89.3\%$ for $x=1.0$ \cite{fufeng}. As $x$ diminishes, $f_R$ decreases considerably. This point can be understood by considering the case of $\epsilon\rightarrow 0$. In such a case, as demonstrated in Appendix~\ref{App:A}, due to the concavity of $f_R$ as a function of $\mathcal{R}_0$, the decrease of epidemic size $f_{Ra}$ in group $A$ cannot be offset by the increase of $f_{Rb}$ in group $B$ and consequently the final epidemic size of the whole system will decrease continuously as $x$ decreases. In particular, when the value of $x$ is less than $0.25$, the value of $R_{0a}$ will be smaller than unity, which means that the epidemic cannot spread throughout group $A$. Hence $f_R$ of the whole population is mainly contributed by $f_{Rb}$ and converges approximately to a value $\approx0.5$ for $x<0.25$. With the increment of $\epsilon$, the more frequent contact between the two groups will infect more individuals in group $A$, while the somewhat less frequent contact among those individuals from group $B$ has just a slight impact on the final $f_{Rb}$ (see Appendix~\ref{App:A}). The introduction of heterogeneity of the infection rate can greatly suppress the prevalence of the infectious disease.

\begin{figure}
\epsfig{figure=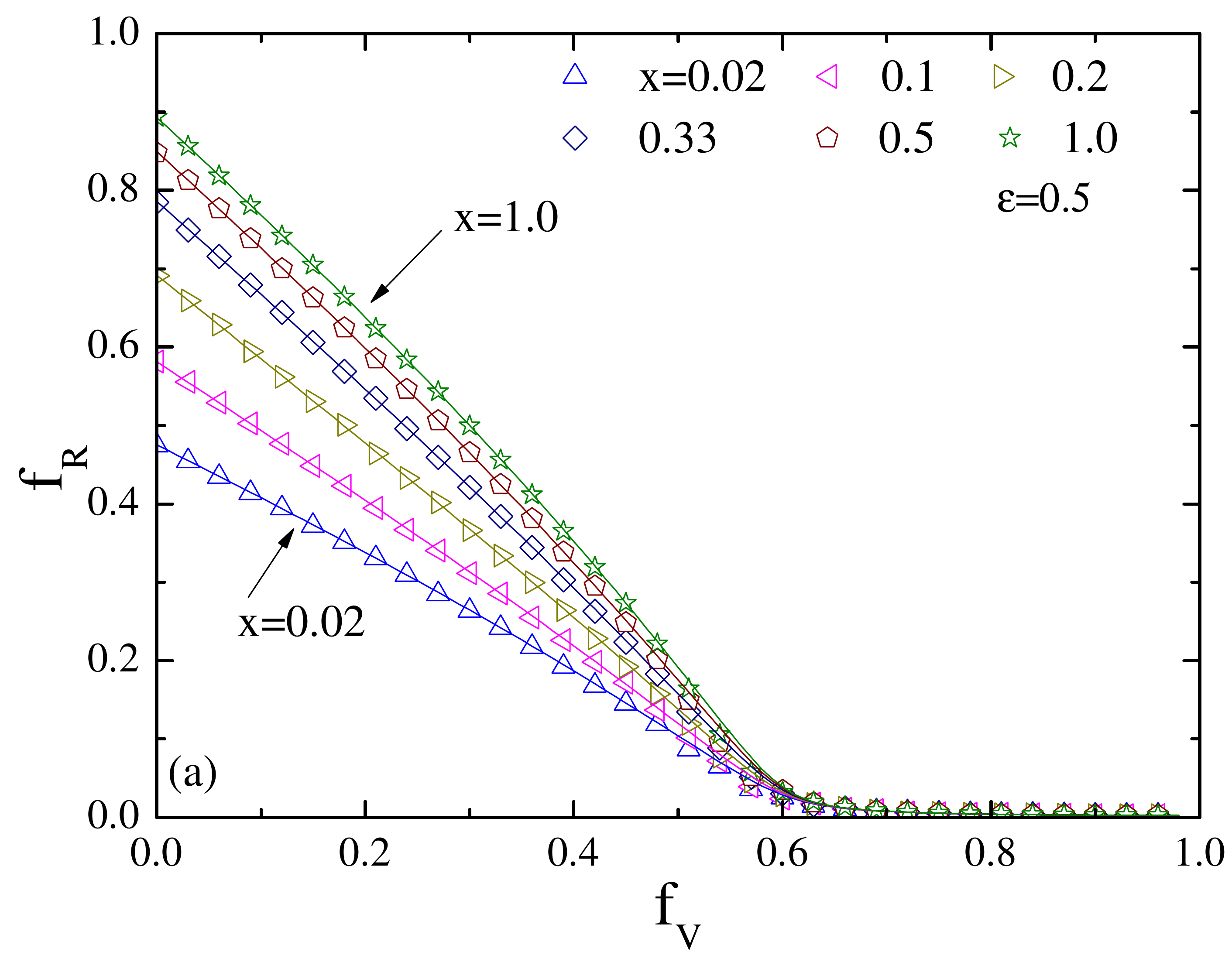,width=\linewidth}
\epsfig{figure=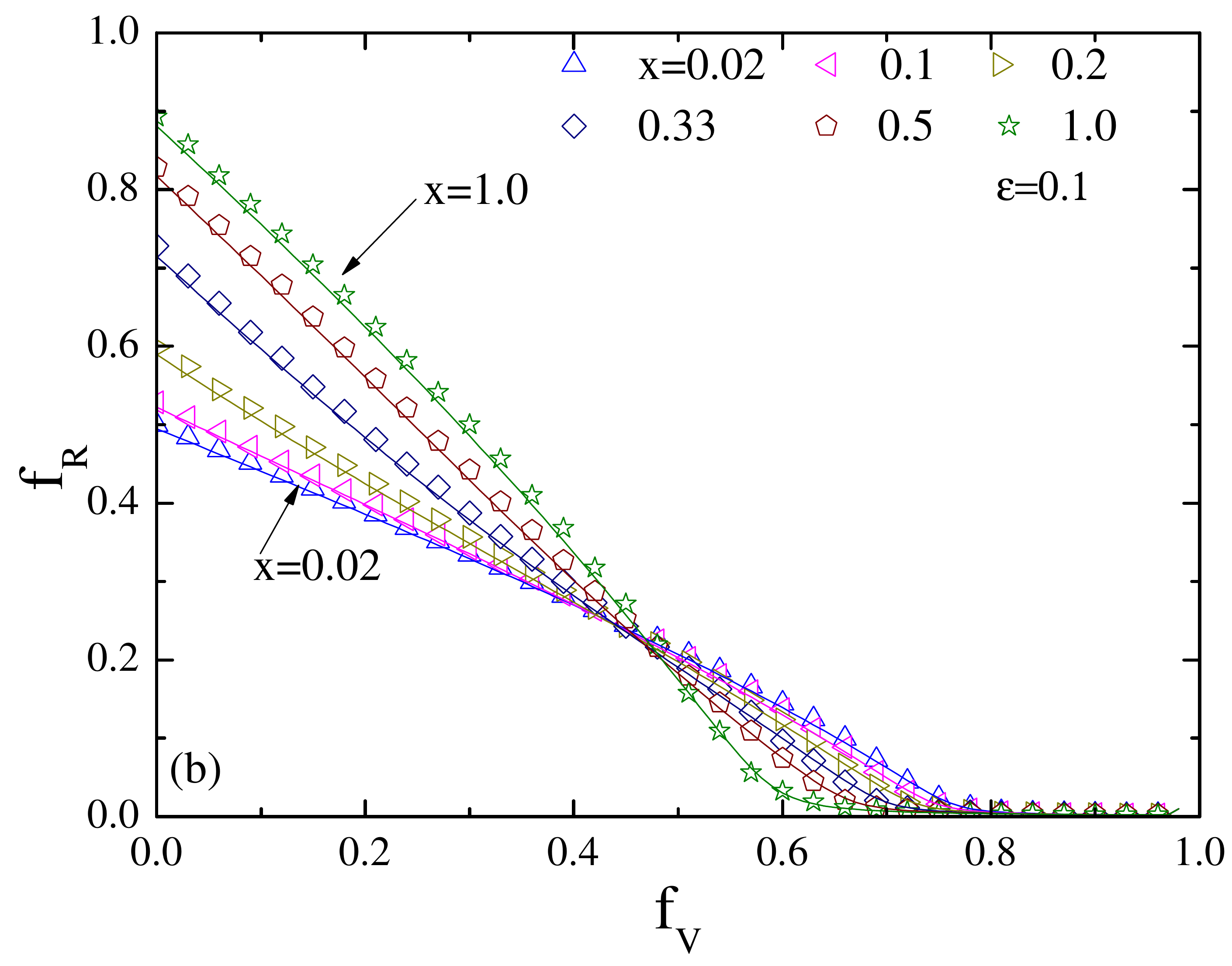,width=\linewidth}
\caption{(Color online) Epidemic spreading in the metapopulation model with the vaccinated compartment. The final epidemic size $f_R$ is plotted as a function of the fraction of vaccinated individuals $f_V$ for several different values of the relative disease transmission rate $x$. The lines are for analytical predictions from deterministic equations and the symbols are obtained by simulations. The cross contact coefficient (a) $\epsilon=0.5$ and (b) $\epsilon=0.1$. Other parameters are the same as in Fig.~\ref{Fig:meta-without}. Simulation results are averaged over 100 independent runs.}
\label{Fig:meta-with}
\end{figure}

\subsection{Metapopulation with vaccinated compartment}
We now incorporate the vaccinated compartment into the epidemic spreading in the metapopulation model. We denote by $f_{Va}$ the proportion of the population initially vaccinated in group $A$. In our work we assume the same fraction of initially vaccinated individuals for the two groups, that is, $f_{Va}=f_{Vb}=f_V$. For given values of $f_V$, $x$, and $\epsilon$, we obtain the final epidemic size by implementing stochastic simulations as described in Sec.~\ref{model}. The simulation results are summarized in Fig.~\ref{Fig:meta-with}, which are in good agreement with those predicted by numerically solving Eqs.~(\ref{dSdt})--(\ref{dRdt}).

The overall result is that with the involvement of the vaccinated compartment, the final epidemic size will gradually decrease with the increase of $f_V$, which is expected since vaccination can provide perfect immunity to the infectious disease and a sufficiently large fraction of vaccinated individuals can completely prohibit the propagation of the infectious disease. Though the difference between $f_R$ for $x=1.0$ and that for $x<1.0$ is vanishing in the limit of large $f_V$, there exists a qualitative difference for the variation. When the individuals from the two groups interact quite frequently $\epsilon=0.5$, the smaller the relative disease transmission rate $x$ is, the smaller the final epidemic size $f_R$ is. Such a dynamic scenario, however, changes when the interaction frequency among the individuals from distinct groups is decreased. Specifically, a crossover behavior of $f_R$ as a function of $f_V$ emerges as the parameter $\epsilon$ drops close to zero. We notice that there arises a critical value of $f_V$, say, $f_{Vc}$ (whose value is about 0.45), below which the presence of heterogeneity in infection rate for the individuals from different groups can hinder the epidemic spreading, while above which the opposite effect takes place (see Appendix~\ref{App:B} for more details). It is worth pointing out that for sufficiently small $\epsilon$, the individuals in the two groups almost interact with others within the same group, which leads to the clustering of susceptible individuals with a high infection rate of the disease (in group $B$). Consequently, the disease prevalence is enlarged as compared to the case of a homogeneous interaction pattern of the two groups [e.g., the curve for $x=0.02$ in the case of $\epsilon=0.1$ is always above that in the case of $\epsilon=0.5$ (not shown here)].

\begin{figure}
\begin{align*}
\epsfig{figure=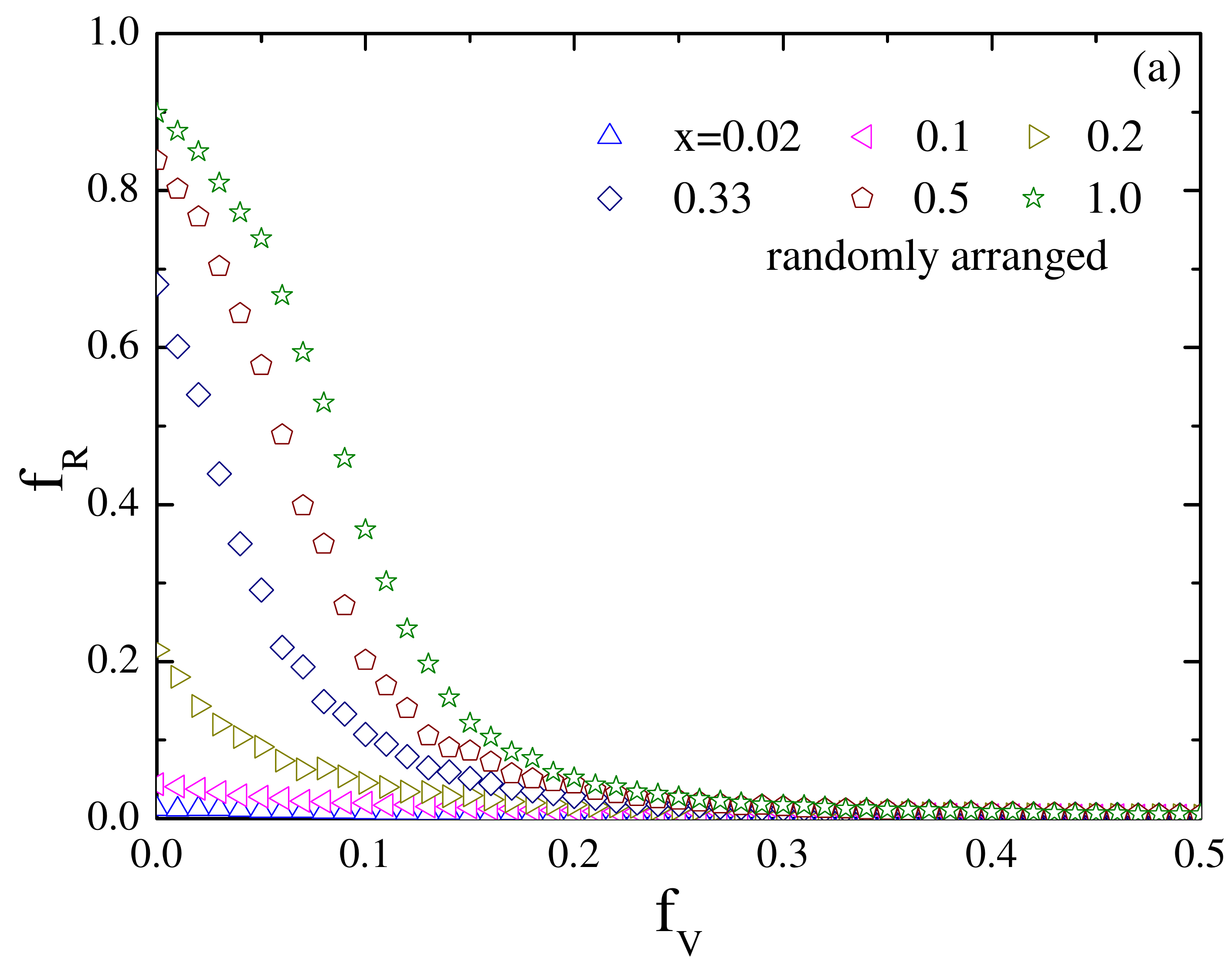,width=\linewidth}\\
\epsfig{figure=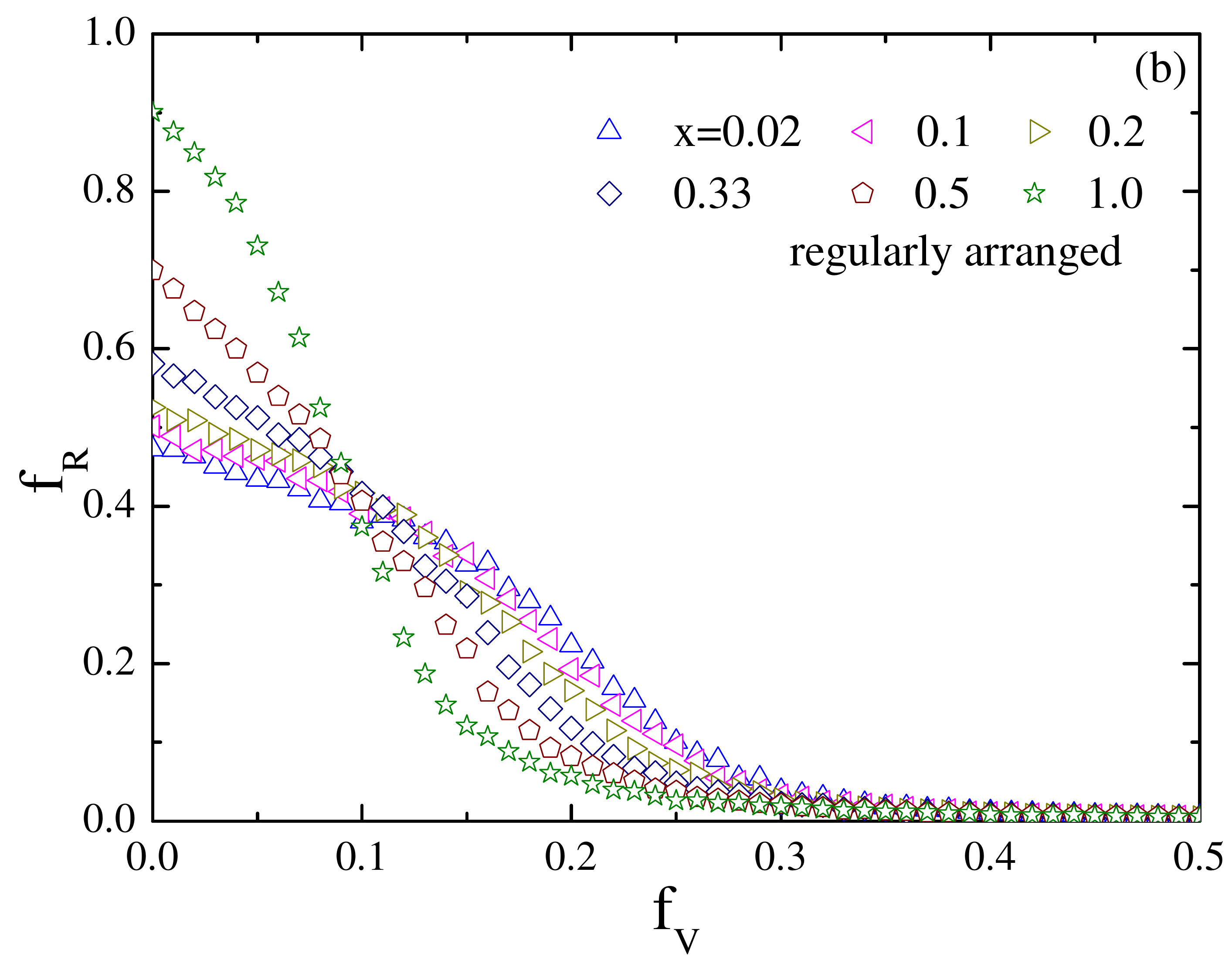,width=\linewidth}
\end{align*}
\caption{(Color online) Epidemic spreading in spatially structured populations with the vaccinated compartment. The final epidemic size $f_R$ is plotted as a function of the vaccination level $f_V$ for several different values of the relative disease transmission rate $x$. The different types of individuals are (a) randomly arranged as in Fig.~\ref{Fig:lattice}(b) and (b) regularly arranged as in Fig.~\ref{Fig:lattice}(c). The parameters are the total population size $N$=100$\times$100, average value of the disease transmission rate $\langle r\rangle=0.46$ day$^{-1}$ person$^{-1}$, recovery rate $g=\frac{1}{3}$ day$^{-1}$, and number of initial infection seeds $I_a$=$I_b$=10. Simulation results are averaged over 100 independent runs.}
\label{Fig:spatial}
\end{figure}

\begin{figure*}
\begin{align*}
\epsfig{figure=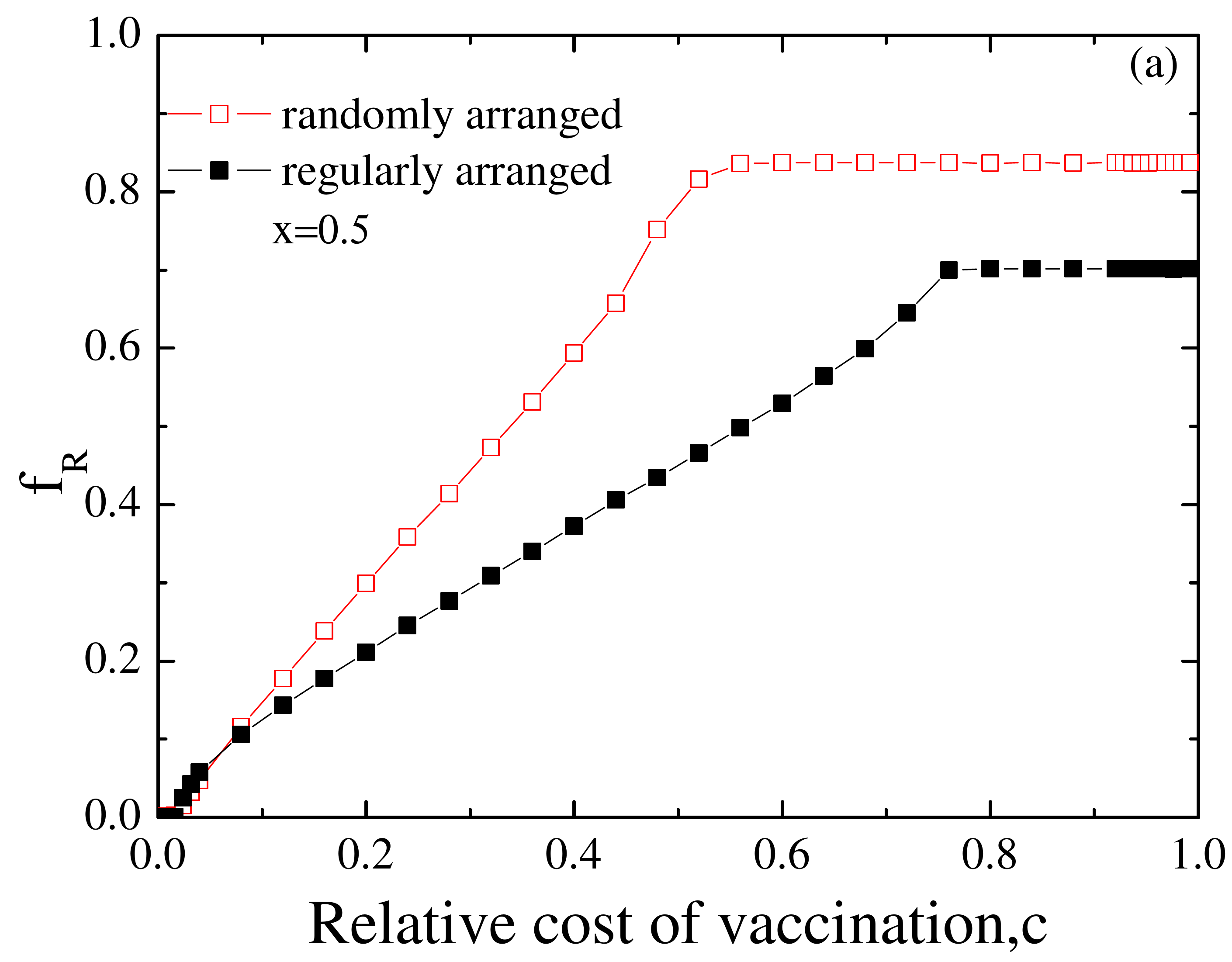,width=0.33\linewidth}
\epsfig{figure=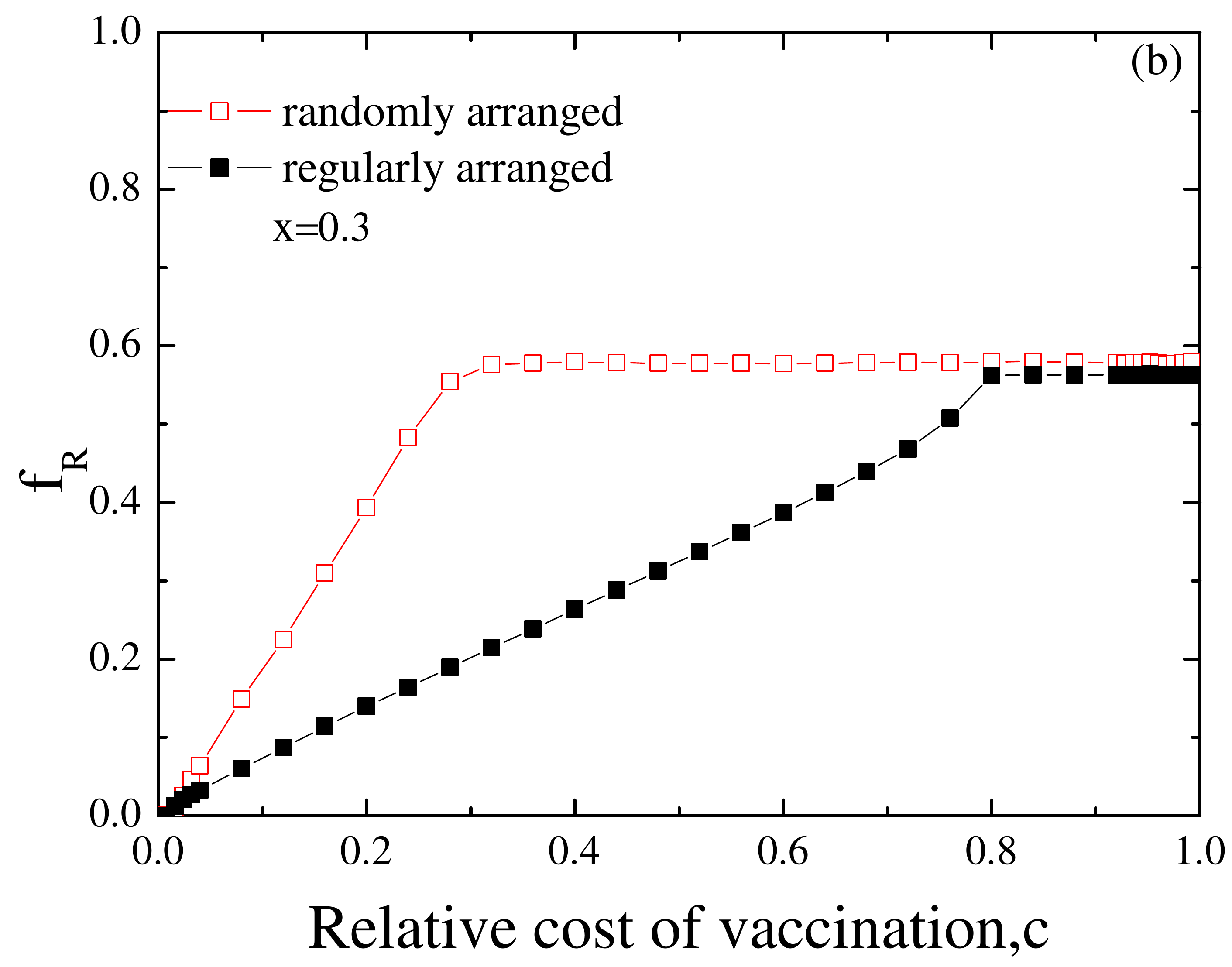,width=0.33\linewidth}
\epsfig{figure=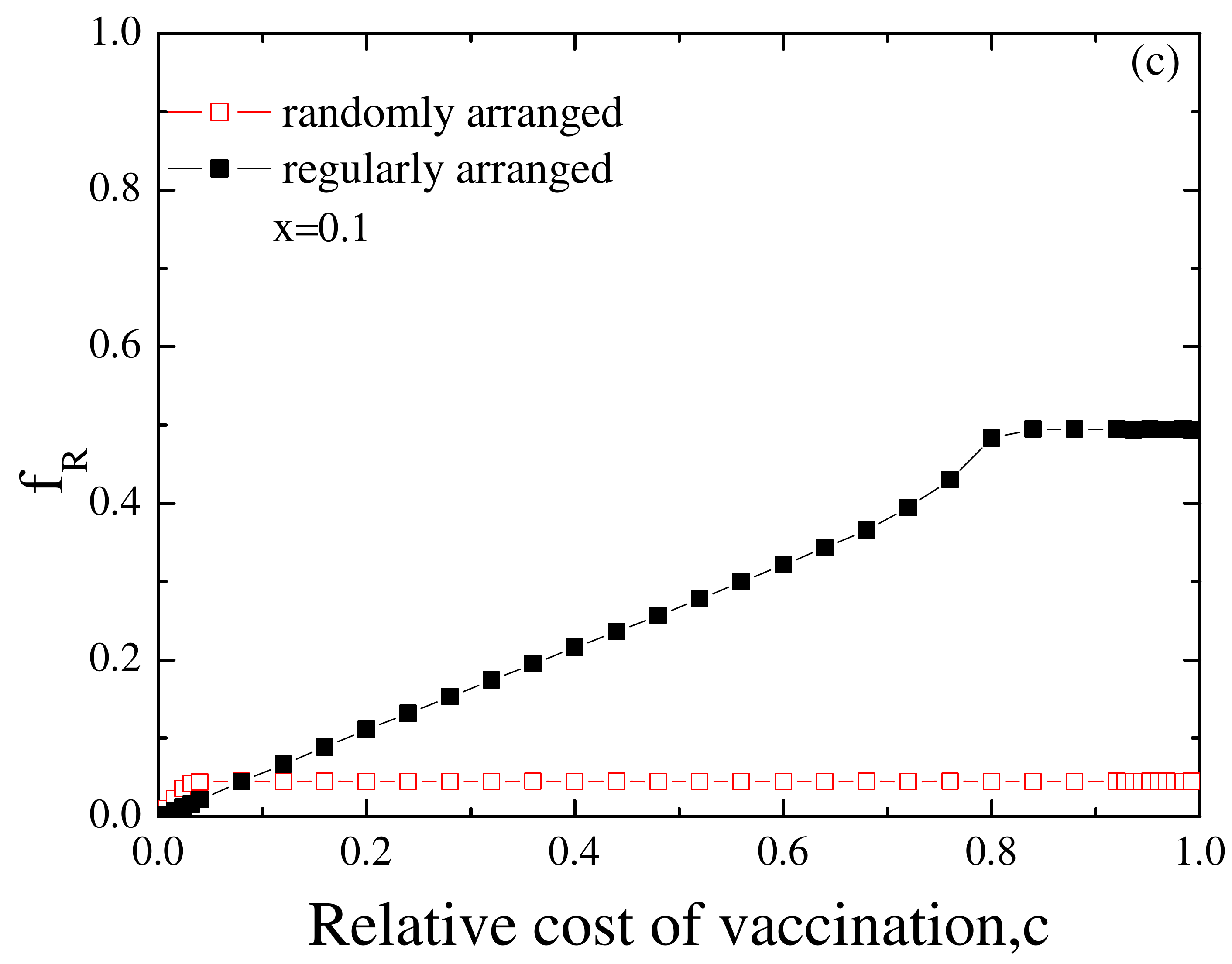,width=0.33\linewidth}
\end{align*}
\begin{align*}
\epsfig{figure=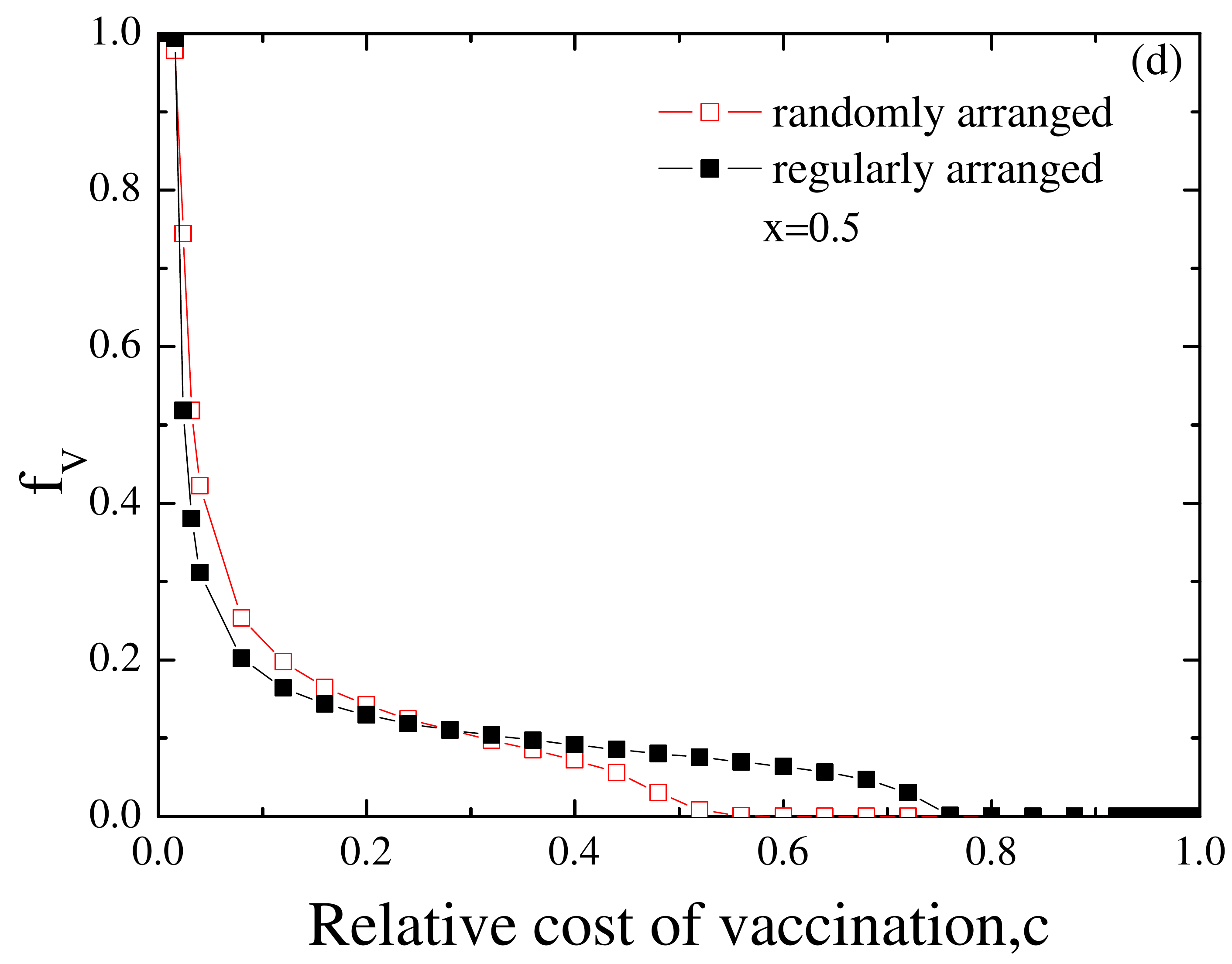,width=0.33\linewidth}
\epsfig{figure=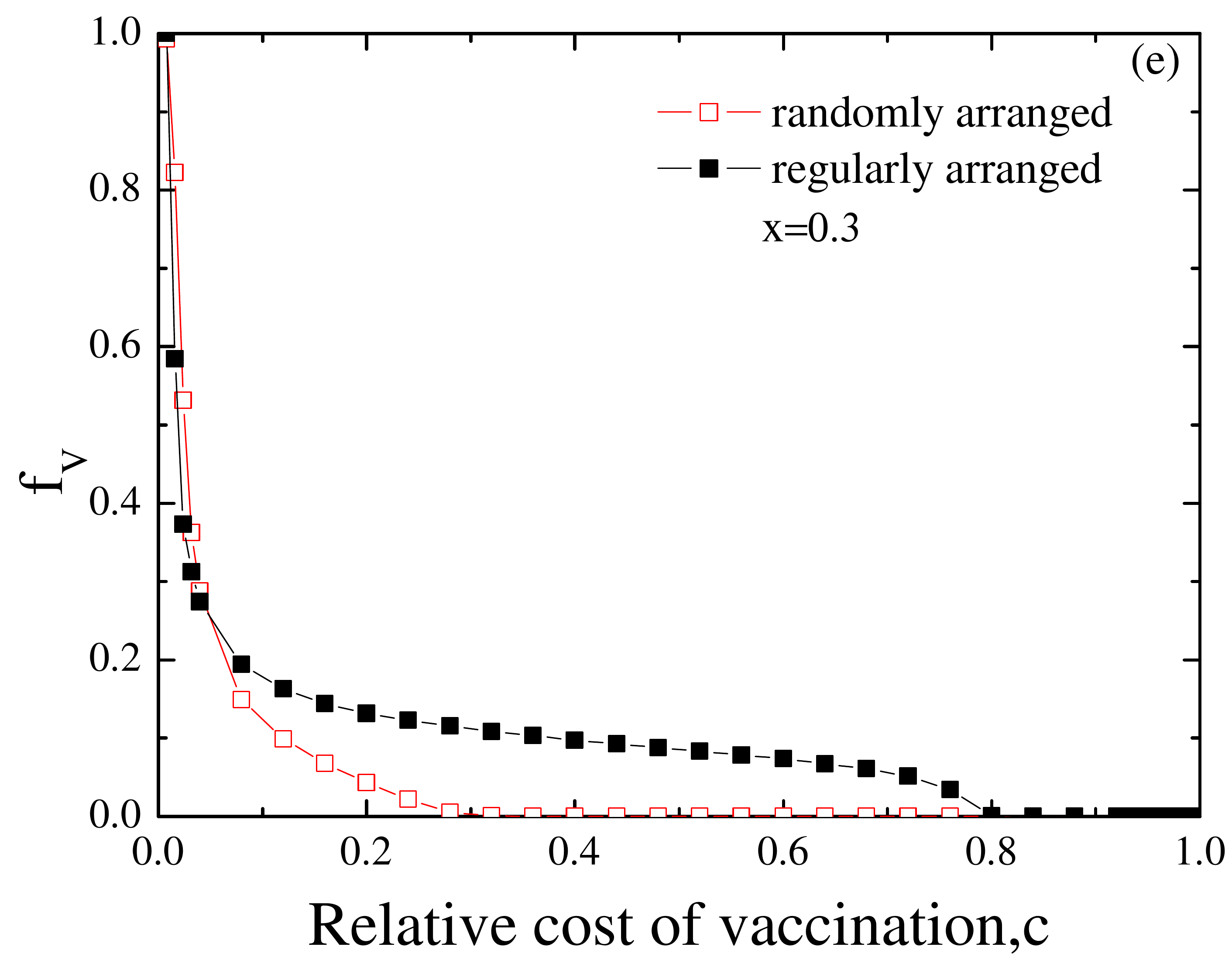,width=0.33\linewidth}
\epsfig{figure=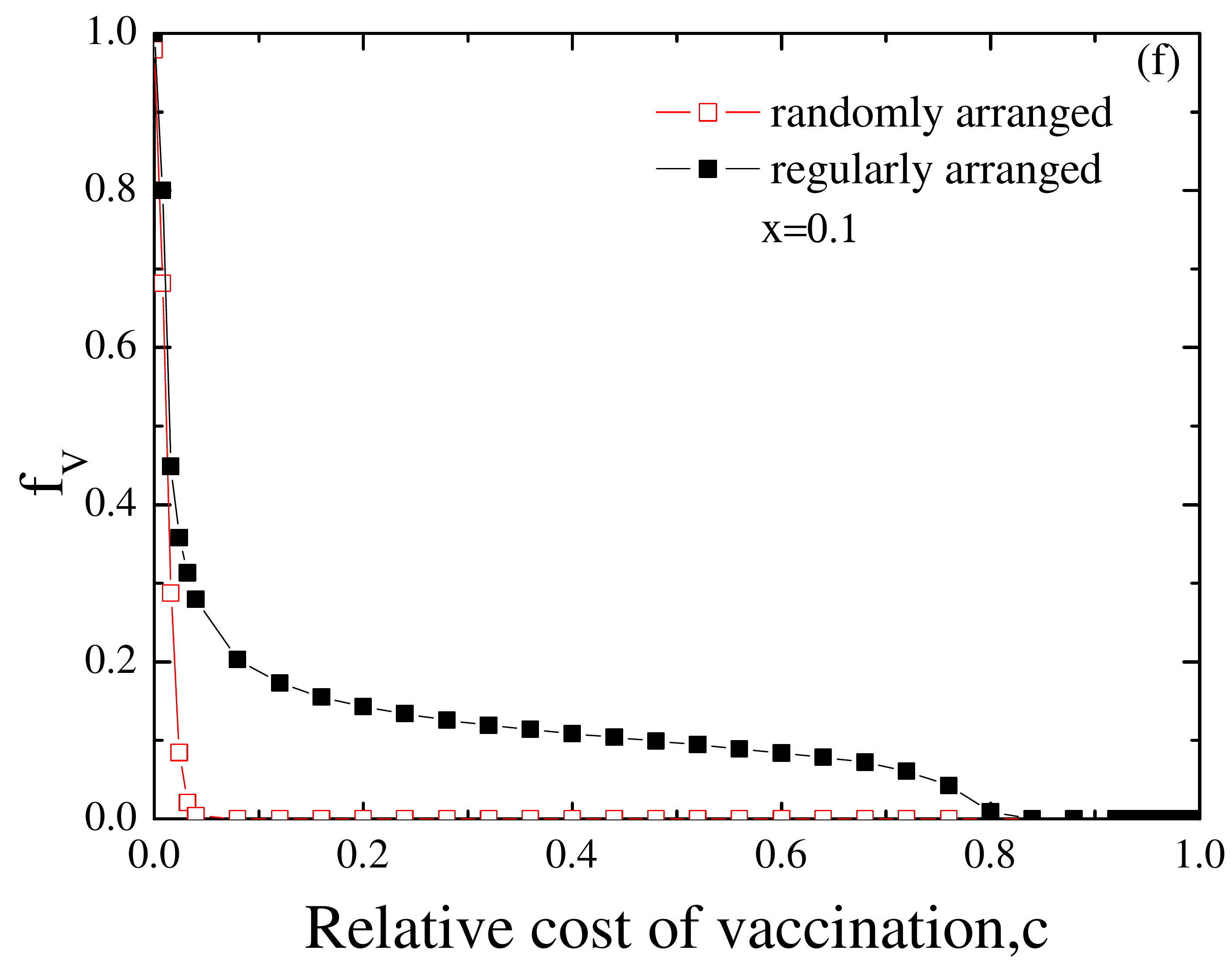,width=0.33\linewidth}
\end{align*}
\caption{(Color online) Epidemic spreading and vaccination dynamics in spatially structured populations. (a)--(c) The final epidemic size $f_R$ and (d)--(f) the final vaccination coverage $f_V$ are plotted as a function of the cost for vaccination $c$ for three typical values of the relative disease transmission rate $x$. Open and closed symbols correspond to the results yielded for randomly and regularly arranged populations, respectively. Other parameters are the same as in Fig.~\ref{Fig:spatial}. Simulation results are averaged over 100 independent runs. The lines are a guide to the eyes.
}\label{Fig:game}
\end{figure*}

\subsection{Spatially structured population with vaccinated compartment}
Now we study our model in a spatially structured population, where the individuals are located on a square lattice. For the sake of comparison, we calibrated the epidemic parameters to ensure that the infection risk in an unvaccinated population (without variation of infection) is equal across all population structures, that is, $f_R$ for $x=1$ in the case of spatially structured population should be the same as $f_R$ for $x=1$ in the case of a metapopulation. The simulation results are displayed in Fig.~\ref{Fig:spatial}, from which we note that the final epidemic size $f_R$ decreases much more rapidly as compared to that in the metapopulation case when the vaccination level increases. When the two types of individuals are \emph{randomly} prearranged, $f_R$ decreases monotonically as the variation of infection increases for each vaccination level. Noticeably, we find that the crossover behavior of $f_R$ as a function of $f_V$ still exists when the interaction frequency between the two types of individuals reduces to a very low level. From Fig.~\ref{Fig:spatial}(b) we can see clearly that there is a crossing point near $f_{Vc}=0.1$. For $f_V<f_{Vc}$, the heterogeneity in infection can efficiently hinder the disease spreading, while it promotes the propagation for $f_V>f_{Vc}$, similar to the results in Fig.~\ref{Fig:meta-with}(b) obtained for the metapopulation model.

\subsection{Spatially structured population with vaccination dynamics}
In what follows we investigate how the vaccination dynamics (i.e., we allow the individuals to change their vaccination behavior based on previous experience~\cite{fufeng}) affects the epidemic spreading in structured populations. In the initial state, we randomly choose half of the population to get vaccinated. At the end of each epidemic spreading season, we give the individuals a chance to update their strategies for vaccination before the new one starts. We implement a pairwise comparison process for the strategy updating. Specifically, whenever an individual $i$ updates one's vaccination strategy, one just chooses an individual $j$ randomly from one's nearest neighbors to compare their cost (or payoff) and then adopts the vaccination choice of $j$ with a probability dependent on the payoff difference~\cite{Szab,Arne,Traulsen}
\begin{equation}\label{fai}
q_{ij}=\frac{1}{1+\exp[-\beta(P_j-P_i)]},
\end{equation}
where $P_i$ and $P_j$ correspond to the payoffs of the two involved individuals and $\beta$ denotes the strength of the selection. Unless otherwise specified, we select $\beta=1.0$, implying that better-performing individuals are readily imitated, but it is not impossible to adopt the behavior of an individual performing worse. What we are interested in this case is how many individuals are infected and the vaccination coverage in the final stable state. The results shown in Fig.~\ref{Fig:game} are the average of the last 1000 iterations among the total 5000 in 100 independent simulations.

We plot in Fig.~\ref{Fig:game} the epidemic size $f_R$ and the vaccination level $f_V$ in the steady state as a function of the relative cost for vaccination $c$ for two differently arranged populations on square lattice. From Figs.~\ref{Fig:game}(a)--\ref{Fig:game}(c) we observe that as the value of $x$ goes down, i.e., the heterogeneity in infection rate for the two types of individuals becomes more notable, the final epidemic level in the randomly arranged population (the open symbols) changes much more evidently than that in the case of regularly arranged population (the closed symbols). In particular, for $x=0.5$, the final $f_R$ in the randomly arranged population is always greater than that in the regularly arranged population as $c$ increases, albeit the vaccination level in the former case is slightly larger than that in the latter case for $c\lesssim0.25$ [see Fig.~\ref{Fig:game}(d)]. For $x=0.3$, in the randomly arranged population, though the growth trend of $f_R$ is more apparently for small $c$, it attains at a smaller level for large enough values of $c$ (when the vaccination level evolves to zero), which is comparable to the case of a regularly arranged population. As $x$ decreases even to 0.1, $f_R$ in a randomly arranged population can just grow to a much lower level as compared to that in the case of a regularly arranged population, despite the fact that the vaccination level is zero for most $c$ values [see Fig.~\ref{Fig:game}(f)]. The reason is that the $A$-type individuals are difficult to infect even though they did not receive a vaccine when $x$ is too small and as such they play the role of a \emph{natural obstructer} to prevent large-scale spreading of the disease in the population. In addition, those unvaccinated $A$-type individuals will attract other individuals to not get vaccinated, giving rising to very low level of vaccination in the steady state [Figs.~\ref{Fig:game}(e) and \ref{Fig:game}(f)]. For a regularly arranged population, however, since the $B$-type individuals are clustered together, they are very prone to the attack of disease, and consequently the final epidemic can reach a rather large level.

\section{Conclusion and Discussion}~\label{conclusion}

In summary, we have incorporated the heterogeneity in infection rate of individuals and also the vaccination dynamics into the traditional susceptible-infected-recovered compartmental epidemic model to study their potential effects on the disease prevalence and vaccination coverage. For this purpose, we have considered a more practical framework where the whole population is classed into two types of groups whose members are endowed with different capabilities in catching a disease. To keep things simple, the individuals within the same group are assumed to be identical in their infection rate. The proposed model has been investigated in a simple metapopulation and spatially structured populations, with and without involvement of vaccination, by using numerical simulations as well as analytical treatments.

We have shown that whether the introduction of heterogeneity in the infection rate of the individuals exerts positive or negative effects (i.e., hampers or expedites) on the epidemic spreading depends closely on both the extent of the heterogeneity of the disease transmission rate and the interaction frequency among the individuals from different groups. To be more specific, the heterogeneity in infection rate can always give rise to a decrease of the final epidemic size provided the individuals from different groups interact with equal likelihood. Nonetheless, as the individuals become more inclined to interact mainly with others from the same group, the heterogeneity in infection rate can hinder the epidemic spreading only in the situation that the fraction of individuals vaccinated is low enough. Very surprising, this just facilitates the epidemic spreading in a regime with the presence of a large fraction of vaccinated individuals (but not large enough to eradicate the disease completely).

Our work is expected to provide some valuable instructions for the prediction and intervention of epidemic spreading in the real world. The results summarized in Figs.~\ref{Fig:meta-without}--\ref{Fig:game} suggest that when evaluating the seriousness of an epidemic, we should take into account both the factors of the diversity of the infection rate of the individuals and the interaction patterns among them simultaneously; otherwise we may overestimate or underestimate the spreading trend. Alternatively, without such considerations, we may overshoot or undershoot the desired amount of action when developing, regulating, and making vaccine policy. In addition, when individuals are allowed to change their vaccination decisions according to their experience and observations, we find that as the heterogeneity in infection rate for the two types of individuals becomes more noticeable, the final epidemic level in randomly arranged population changes much more evidently than that in the case of a regularly arranged population, hence giving us a vital clue as to how to make efficient vaccine campaign, namely, we should distribute the vaccine in the population as widely as possible so that the spreading path of the disease can be efficiently suppressed.

To summarize, our proposed model captures essential elements in real-world epidemic spreading, which has not been fully discussed previously. Therefore, we believe our results will give some insights to the policy makers. There are still many issues, such as diversity of recovery rate, heterogeneous cost for infection and vaccination, and more complex contact-network structures, which are totally overlooked in the present work and deserve to be explored in the future. In addition, the spread of awareness of the epidemic and/or the vaccination sentiment would also impact greatly the vaccination behavior of the individuals and hence the epidemic outbreaks~\cite{Funk2009pnas,Ruan2012pre,Campbell2013sr}. We hope our work could stimulate further work in this line of research.

\acknowledgments
This work was supported by the National Natural Science Foundation of China (Grants No. 11005051, No. 11005052, and No. 11135001).

\section{Appendix A}\label{App:A}
Here we present theoretical analysis for the simple metapopulation model. For convenience, let us denote $\langle r\rangle N/g$ by $C$, which is kept as a constant. The combination of Eqs.~(\ref{dS_xdt})--(\ref{dR_xdt}) with Eq.~(\ref{r_a}) yields
\begin{eqnarray}
&\frac{dS_a}{dR_a}=-\frac{2x}{1+x}S_aC\left[\left(1-\epsilon\right)+\epsilon \frac{I_b}{I_a}\right]\label{b1},\\
&\frac{dS_a}{dR_b}=-\frac{2x}{1+x}S_aC\left[\left(1-\epsilon\right) \frac{I_a}{I_b}+\epsilon \right]\label{b2},\\
&\frac{dS_b}{dR_b}=-\frac{2}{1+x}S_bC\left[\left(1-\epsilon\right)+\epsilon \frac{I_a}{I_b}\right]\label{b3},\\
&\frac{dS_b}{dR_a}=-\frac{2}{1+x}S_aC\left[\left(1-\epsilon\right) \frac{I_b}{I_a}+\epsilon\right]\label{b4}.
\end{eqnarray}
After eliminating $\frac{I_a}{I_b}$ and $\frac{I_b}{I_a}$ from these equations we readily obtain
\begin{eqnarray}
&\frac{\epsilon}{S_a}dS_a-\frac{x(1-\epsilon)}{S_b}dS_b= \frac{2Cx(1-2\epsilon)}{1+x}dR_b\label{b5},\\
&\frac{1-\epsilon}{S_a}dS_a-\frac{x\epsilon}{S_b}dS_b= \frac{2Cx(1-2\epsilon)}{1+x}dR_a\label{b6}.
\end{eqnarray}
Now we integrate these two equations with respect to time from $0$ to $\infty$. By using the initial condition $S_a(0)=S_b(0)\approx 1$ and $R_a(0)=R_b(0)=0$ and the final state $I_a(\infty)=I_b(\infty)=0$, we get the following two transcendental equations for the final epidemic size $R_a(\infty)$ and $R_b(\infty)$ for each group:
\begin{eqnarray}
&\ln\left[1-R_a(\infty)\right]=\frac{2Cx}{1+x}\left[(1-\epsilon)R_a(\infty)+\epsilon R_b(\infty)\right]\label{b7},\\
&\ln\left[1-R_b(\infty)\right]=\frac{2C}{1+x}\left[(1-\epsilon)R_b(\infty)+\epsilon R_a(\infty)\right]\label{b8}.
\end{eqnarray}
What we want to figure out is the relationship between the final epidemic size $f_R$ and the cross coefficient $\epsilon$, so we take a derivative of Eqs.~(\ref{b7}) and~(\ref{b8}) with respect to $\epsilon$ and get

\begin{eqnarray}
\nonumber&\frac{1}{R_a(\infty)-1}\frac{dR_a(\infty)}{d\epsilon}\label{b9}\\
&=-\frac{2Cx}{1+x}\left[-R_a(\infty)+(1-\epsilon)\frac{dR_a(\infty)}{d\epsilon}\label{b9}
+R_b(\infty)+\epsilon \frac{dR_b(\infty)}{d\epsilon}\right]\label{b9},
\end{eqnarray}
\begin{eqnarray}
\nonumber&\frac{1}{R_b(\infty)-1}\frac{dR_b(\infty)}{d\epsilon}\\
&=-\frac{2Cx}{1+x}\left[-R_b(\infty)+(1-\epsilon)\frac{dR_b(\infty)}{d\epsilon}
+R_a(\infty)+\epsilon \frac{dR_a(\infty)}{d\epsilon}\right]\label{b10}.
\end{eqnarray}
After doing some algebra we obtain
\begin{eqnarray}
\nonumber&\left(\frac{1}{1-R_a(\infty)}-\frac{2Cx}{1+x}\right)\frac{dR_a(\infty)}{d\epsilon}\\
&=-\left(\frac{x}{1-R_b(\infty)}-\frac{2Cx}{1+x}\right)\frac{dR_b(\infty)}{d\epsilon}\label{b11}.
\end{eqnarray}

\begin{figure}
\epsfig{figure=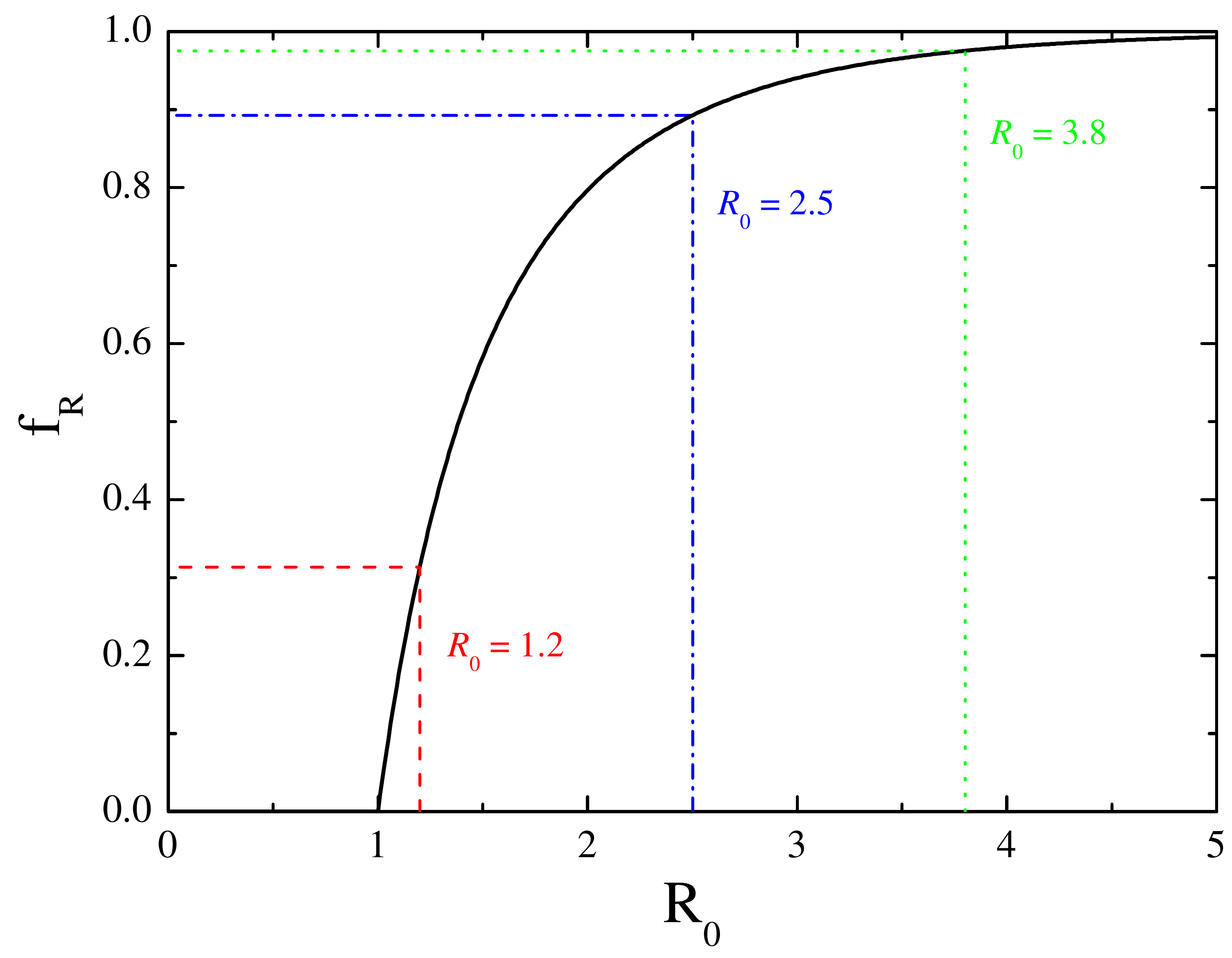,width=0.87\linewidth}
\caption{(Color online) Solutions of the equation $R(\infty)=1-\exp^{-R_0R(\infty)}$, where the final epidemic size $f_R=R(\infty)$ as a function of basic reproduction ratio $R_0$ is shown. }
\label{Fig:A2}
\end{figure}

We can rewrite this equation as
\begin{equation}\label{b12}
\frac{dR_a(\infty)}{d\epsilon}=-K\frac{dR_b(\infty)}{d\epsilon},
\end{equation}
where \begin{equation}\label{K}
K=\frac{\frac{x}{1-R_b(\infty)}-\frac{2Cx}{1+x}} {\frac{1}{1-R_a(\infty)}-\frac{2Cx}{1+x}}.
\end{equation}
In the case of $\epsilon=0$, it is easy to verify numerically that $K>1$ for all our $x$  values of interest (say, $x>0.01$). More intuitively, for SIR model in a well-mixed population, the final epidemic size is determined by the self-consistent equation $R(\infty)=1-\exp^{-R_0R(\infty)}$. Figure~\ref{Fig:A2} features the solutions, from which we note that $f_R$ is a concave function of $R_0$. If we decrease the value of $x$ such that in the limit of $\epsilon=0$ the variables $R_{0a}$ and $R_{0b}$ always satisfy the relationships $R_{0a}<R_{0b}$ and $\left(R_{0a}+R_{0b}\right)/2=C=R_0$, then due to the concave curvature, the variation of the final epidemic size in group $A$ will be more remarkable than that in group $B$, as illustrated in Fig.~\ref{Fig:A2}. For each fixed value of $x$, as $\epsilon$ increases, the increasingly frequent contact among the individuals from different groups will have a greater affect on $R_a(\infty)$ than on $R_b(\infty)$ (as long as $x$ is not too small), giving rise to the increase of $f_R$ in the whole population. Since $R_a(\infty)$ increases and $R_b(\infty)$ decreases with the increment of $\epsilon$, the value of $K$ will decrease monotonically according to Eq.~(\ref{K}), which is reflected correctly in Fig.~\ref{Fig:meta-without}.

\section{Appendix B}\label{App:B}
Here we demonstrate the existence of the crossover behavior for the curves of $f_R$ as a function of $f_V=y$ for different values of $x$. In a well-mixed population, we know the final fraction of recovered population for the SIR model satisfying the equation $R(\infty)=1-\exp^{-R_0R(\infty)}$. When a proportion $y$ of preemptive vaccination in introduced before the epidemic starts, we can readily obtain $R(\infty)=(1-y)\left(1-\exp^{-R_0R(\infty)}\right).$
For our proposed model, we consider two limited cases. The first case is $x=1$, i.e., the individuals in the two groups are identical, and in such a case we have
\begin{eqnarray}\label{aa}
&R_b(\infty)\mid_{x=1}=(1-y)\left(1-\exp^{-R_{0b}R_b(\infty)\mid_{x=1}}\right),\\
&R(\infty)\mid_{x=1}=R_a(\infty)\mid_{x=1}=R_b(\infty)\mid_{x=1},
\end{eqnarray}
where $R_{0b}=r_bN/g=\langle r\rangle N/g=C$.

\begin{table}
\caption{Comparisons of the intersecting points ($f_{Vc}$, $f_R$) of the curves for $x=1.0$ and $0.02$ predicted by Eqs.~(\ref{a4}), (\ref{aa}), and (\ref{bb}) with those obtained from direct stochastic simulations, with different values of $\epsilon$. \label{tab}}
\begin{ruledtabular}
\begin{tabular}{cccc}
\textrm{$\epsilon=0.05$}&
\textrm{$\epsilon=0.1$}&
\textrm{$\epsilon=0.2$}&
\textrm{$\epsilon=0.3$}\\
\colrule
0.470, 0.237(6)\footnote{Results of ($f_{Vc}$, $f_R$) obtained from stochastic simulations.} & 0.477, 0.228(5) & 0.482, 0.210(2) & 0.507, 0.175(5)\\
0.473, 0.235(7)\footnote{Results of ($f_{Vc}$, $f_R$) predicted by analytical treatments.} & 0.478, 0.227(7) & 0.491, 0.206(4) & 0.512, 0.171(5)\\
\end{tabular}
\end{ruledtabular}
\end{table}

The other limited case is $x\rightarrow0$, which means that the disease transmission rate for the individuals in group $A$ is nearly zero. By approximating $R_a(\infty)\mid_{x\rightarrow0}=0$ and combining Eqs.~(\ref{b7}) with (\ref{b8}) we have
\begin{eqnarray}
&R_b(\infty)\mid_{x\rightarrow0}=(1-y)(1-\exp^{-R_{0b} (1-\epsilon)R_b(\infty)\mid_{x\rightarrow0}})\label{bb},
\end{eqnarray}
\begin{eqnarray}
\nonumber&R(\infty)\mid_{x\rightarrow0}=\frac{R_a(\infty)\mid_{x\rightarrow0}+
R_b(\infty)\mid_{x\rightarrow0}}{2}\\
&= \frac{R_b(\infty)\mid_{x\rightarrow0}}{2},
\end{eqnarray}
where $R_{0b}=r_bN/g=2\langle r\rangle N/g=2C$. We assume that the curves of $f_R$ for the two cases have a crossing point so that
\begin{equation}\label{cc}
R_b(\infty)\mid_{x=1}=\frac{R_b(\infty)\mid_{x\rightarrow0}}{2}.
\end{equation}
Denoting $R_b(\infty)\mid_{x=1}$ by $z$, combining Eqs.~(\ref{aa}), (\ref{bb}), and (\ref{cc}), and recalling that $C=2.5$, we obtain
\begin{eqnarray}
&\exp^{-10(1-\epsilon)z}=2\exp^{-2.5z}-1\label{a4}.
\end{eqnarray}
To validate the assumption, Eq.~(\ref{a4}) must have an exact solution, which means that
\begin{eqnarray}
&\left(\exp^{-10(1-\epsilon)z}\right)^{'}\mid _{z=0}<\left(2\exp ^{-2.5z}-1\right)^{'}\mid _{z=0}.
\end{eqnarray}
Solving the inequality yields $\epsilon<0.5$. That is to say, the crossover behavior will always exist as long as $\epsilon$ is strictly smaller than one-half. From Eqs.~(\ref{a4}) and (\ref{aa}) we have
\begin{equation}\label{a8}
\frac{d\epsilon}{dz}=\frac{1}{10}\left[-\frac{2.5\exp^{-2.5z}} {z(2\exp^{-2.5z}-1)}-\frac{\ln(2\exp^{-2.5z}-1)}{z^2}\right]<0,
\end{equation}
\begin{equation}\label{a9}
\frac{dy}{dz}=\frac{\exp^{-2.5z}+2.5z\exp^{-2.5z}-1}{(\exp^{-2.5z}-1)^2}<0.
\end{equation}
By dividing Eq.~(\ref{a9}) by (\ref{a8}) we get $d\epsilon/dy>0$, which indicates that the crossing point will move to the right (i.e., the curves intersect at larger values of $y=f_V$) with an increase of the cross contact coefficient $\epsilon$. In Table~\ref{tab} we summarize the crossing point values ($f_{Vc}$, $f_R$) of the curves for $x=1.0$ and $0.02$ yielded by the stochastic simulations as well as those predicted by Eqs.~(\ref{a4}), (\ref{aa}), and (\ref{bb}). We notice that the results obtained from different methods match quite well with each other. The invisible differences may be due to the finite-system-size effect. Specifically, with increasing $\epsilon$ the curves for $x=1.0$ and $0.02$ intersect at points with larger $f_{Vc}$.
%\bibliographystyle{apsrev4-1}
%\bibstyle{apsrev4-1}
\bibliography{bbb}
\end{document}